\documentclass{PoS}

\title{Singlet Free Energies and the Renormalized Polyakov Loop in full QCD}

\ShortTitle{Singlet Free Energies}

\author{\speaker{Konstantin Petrov} for RBC-Bielefeld collaboration\\
        Niels Bohr Institute\\
        E-mail: \email{kpetrov@nbi.dk}}

\abstract{
We calculate the free energy of static quark anti-quark pair and the renormalized Polyakov loop
in 2+1- and 3- flavor QCD using $16^3 \times 4$ and $16^3 \times 6$ lattices and improved staggered
p4 action. We also compare the renormalized Polyakov loop with the results of earlier studies.
}

\FullConference{XXIV International Symposium on Lattice Field Theory\\
		 July 23-28 2006\\
		 Tucson Arizona, US}

\begin{document}

\section{Singlet free energies}

We did finite temperature lattice calculations in 3- and (2+1)- flavor QCD in the region of  small quark masses. 
In the case 
of 2+1 flavor QCD we used two light quark masses $m_q=0.1m_s$ and $0.2m_s$ with $m_s$ being the strange quark mass.
In the case of degenerate three flavors the quark masses were roughly $0.15m_s$ and $0.3m_s$. Calculations have be
done on $16^3 \times 4$ and $16^3 \times 6$ lattices. 
The lattice spacing and thus the temperature scale has been fixed using the Sommer scale $r_0=0.469$ fm \cite{gray}.
We used the interpolation Ansatz for the dependence of $r_0$ on the lattice gauge coupling $\beta=6/g^2$ given
in Ref.\cite{us}.
In our simulations we used the exact RHMC algorithm for the
2+1 flavor case while the standard R-algorithm was used in the 3 flavor calculations. 
Further details about our simulations can be found in Refs. \cite{us,michael_lat06}.

On the gauge configurations separated by 50 trajectories we have calculated the singlet free energy of a static
quark anti-quark pair defined as
\begin{equation}
\exp(-F_1(r,T)/T+C)= \frac{1}{3} \langle Tr W(\vec{r}) W^{\dagger}(0) \rangle,
\label{f1def}
\end{equation}
with $W(\vec{r})$ being the temporal Wilson line. The above definition requires gauge fixing and we use the Coulomb gauge 
as this was done in many previous works \cite{ophil02,okacz02,digal03,okaczlat03,okacz04,petrov04,okacz05}.
In the zero temperature limit the singlet free energy defined above coincides with the well known static potential.
In fact the calculations on the static potential in Ref. \cite{milc04} are based on Eq. (\ref{f1def}).  At finite 
temperature $F_1(r,T)$ gives information about in-medium modification of inter-quark forces and color
screening. The singlet free energy $F_1(r,T)$ as well as the zero temperature static potential is 
defined up to additive constant $C$ which depends on the lattice spacing. Since the temperature is varied by changing 
the lattice spacing to compare the free energy calculated at different lattice spacings we normalized it 
at the smallest distance to the following form for the $T=0$ static potential
\begin{equation}
V(r)=-\frac{0.385}{r}+\frac{1.263}{r_0^2} r.
\label{t0pot}
\end{equation}
The above definition gives a very good parameterization of the lattice data for the zero temperature
static potential calculated in Ref. \cite{us}.  

\begin{figure}
\includegraphics[width=8cm]{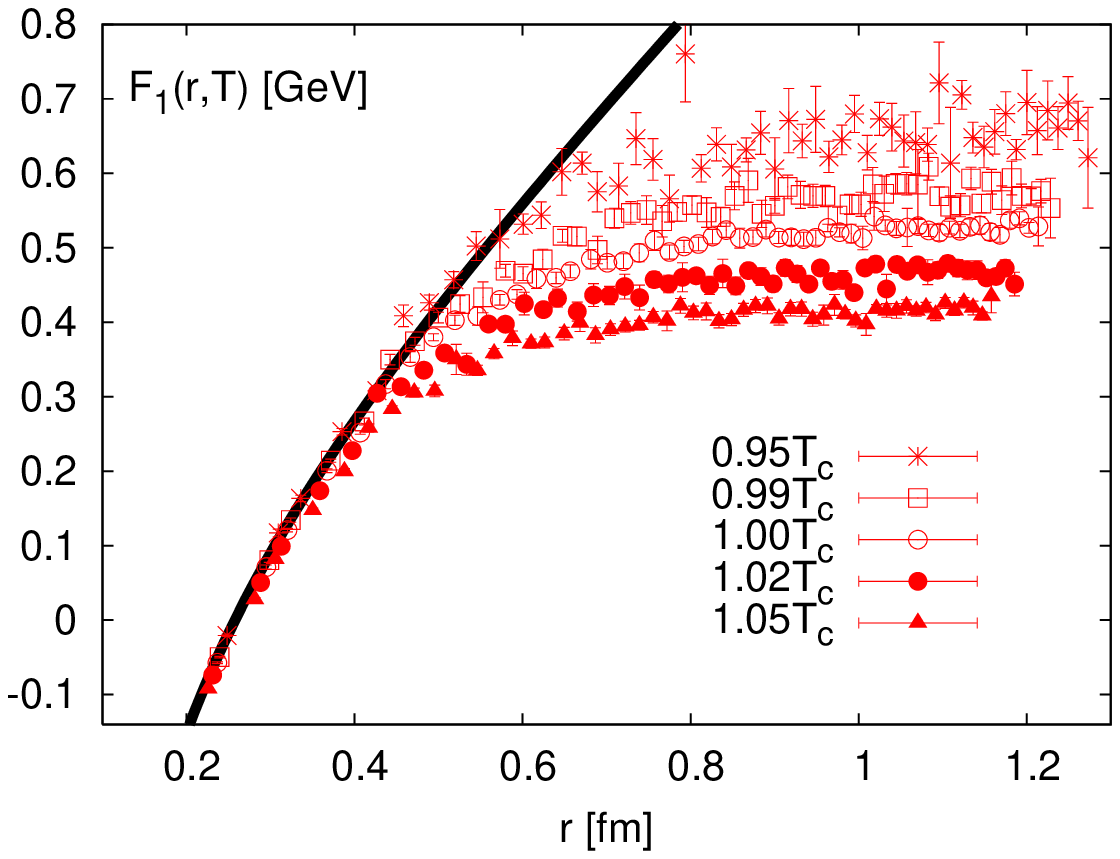}
\includegraphics[width=8cm]{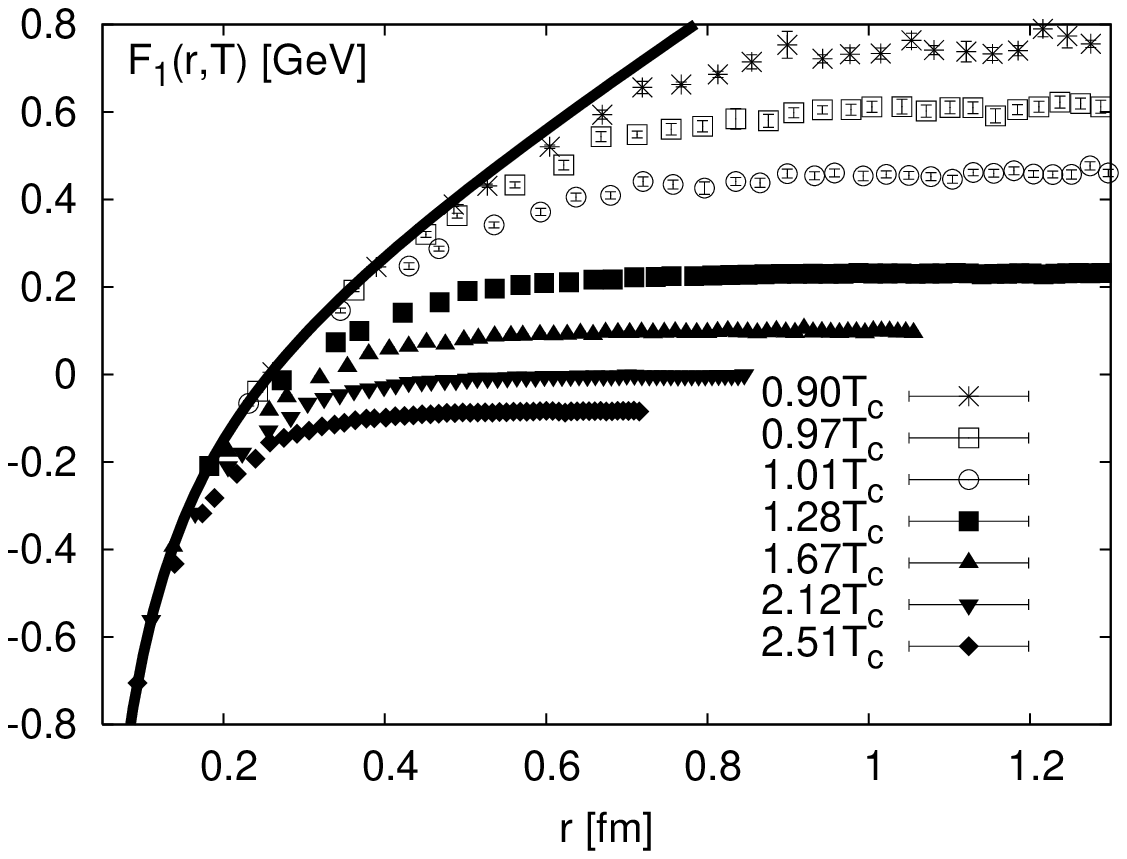}
\caption{
The singlet free energy calculated on $16^3 \times 6$ (left) and on $16^3 \times 4$ lattice for $m_q=0.1m_s$ 
at different temperatures. The thick line shows the parameterization of the zero temperature
potential discussed in the text. 
}
\label{fig:f1}
\end{figure}

\begin{figure}
\centerline{\includegraphics[width=10cm]{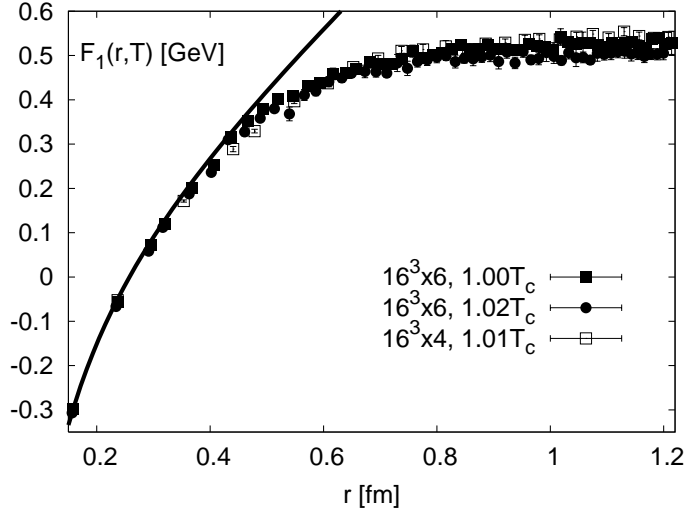}}
\caption{The comparison of the singlet free energy
for $m_q=0.1m_s$ calculated on $16^3 \times 6$ and $16^3 \times 4$ lattices at $T \simeq T_c$. }
\label{fig:f1scaling}
\end{figure}

In Fig. \ref{fig:f1} we show the singlet free energy calculated on $16^3 \times 4$ and $16^3 \times 6$ lattices 
together with the above parameterization of the $T=0$ static potential. As one can see from the figure, 
$F_1(r,T)$ is temperature independent at very small distances and coincides with the zero temperature potential
as expected. At large distances the singlet free energy approaches a constant value. This can be interpreted as
string breaking at low temperature and color screening at high temperatures. Note that the distance where the free
energy effectively flattens off is decreasing with increasing temperatures. This is another indication of color screening.
Although the calculations have been done on quite coarse lattices we see that the results on $F_1(r,T)$ show a fairly good
scaling with the lattice spacing. This is shown in Fig. \ref{fig:f1scaling} where the singlet free energies calculated on $16^3 \times 4$ 
and $16^3 \times 6$ lattices are compared at temperatures $T \simeq T_c$.   

\section{Renormalized Polyakov Loop}

The expectation value of the Polyakov loop $\langle L(\vec{r}) \rangle =\langle Tr W(\vec{r}) \rangle$ 
is the order parameter
for the deconfining transition in pure gauge theories. In full
QCD dynamical quarks  break the relevant $Z(3)$ symmetry explicitly and it is no longer the order parameter. 
Still it remains an interesting quantity to study the deconfinement transition as it shows a rapid increase in the
crossover region \cite{us,milcthermo,fodor06} and can be used to determine the transition temperature \cite{us,fodor06}.
The Polyakov loop defined above strongly depends on the lattice cutoff and has no meaningful continuum limit. On the other hand a correlator of Polyakov loops is a physical quantity and  corresponds to the color averaged free energy up to a normalization constant.
It satisfies the cluster decomposition
\begin{equation}
\exp(-F(r,T)+C)=\frac{1}{9} \langle L(\vec{r}) L^{\dagger}(0) \rangle|_{r \rightarrow \infty}= |\langle L (0)\rangle|^2.
\end{equation}
The normalization constant can be fixed from the color singlet free energy. Moreover, 
at large distances the color singlet free energy and the color averaged free energy approach the same 
constant $F_{\infty}(T)$.
\begin{figure}
\centerline{\includegraphics[width=10cm]{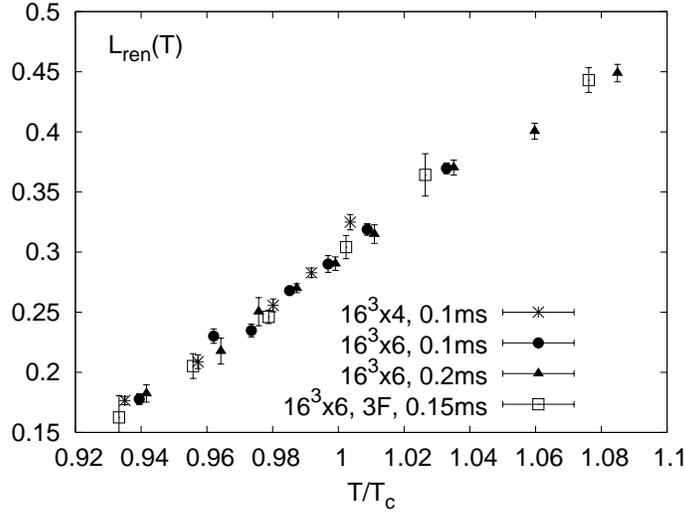}}
\caption{
The renormalized Polyakov in the vicinity of the transition calculated on $16^3 \times 6$ lattices
for different quark masses. Also shown in the figure is the corresponding result from
$16^3 \times 4$ lattice.
\label{fig:lren_sum}
}
\end{figure}
Therefore, following Ref.~\cite{okacz02}, we define the renormalized Polyakov loop as 
\begin{equation}
L_{ren}(T)=\exp(-\frac{F_{\infty}(T)}{2 T}).
\end{equation}
Our numerical results for the renormalized Polyakov loop for different quark masses and two lattice spacings
are summarized in Fig. \ref{fig:lren_sum}. One can see from the figure that $L_{ren}(T)$ shows an almost 
universal behavior as function of $T/T_c$ for all quark masses studied by us, including the case
of three degenerate flavors. This suggests, that in the region of the small quark masses we studied, the
flavor and quark mass dependence of the deconfinement transition can be almost entirely understood in terms
of the flavor and quark mass dependence of the transition temperature $T_c$. Note that the results obtained
on $16^3 \times 4$ lattice are in remarkably good agreement with the results obtained on $16^3 \times 6$ lattices,
indicating again that the cutoff effects are small. 

\begin{figure}
\includegraphics[width=8cm]{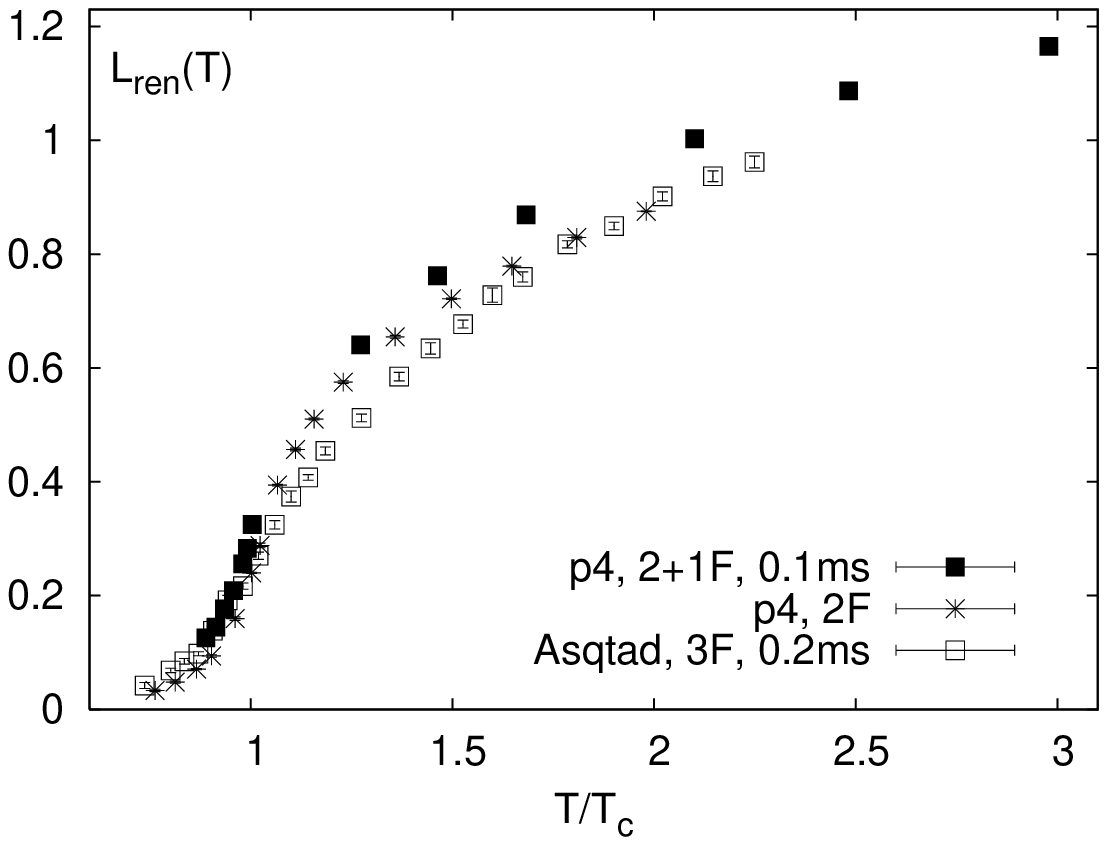}
\includegraphics[width=8cm]{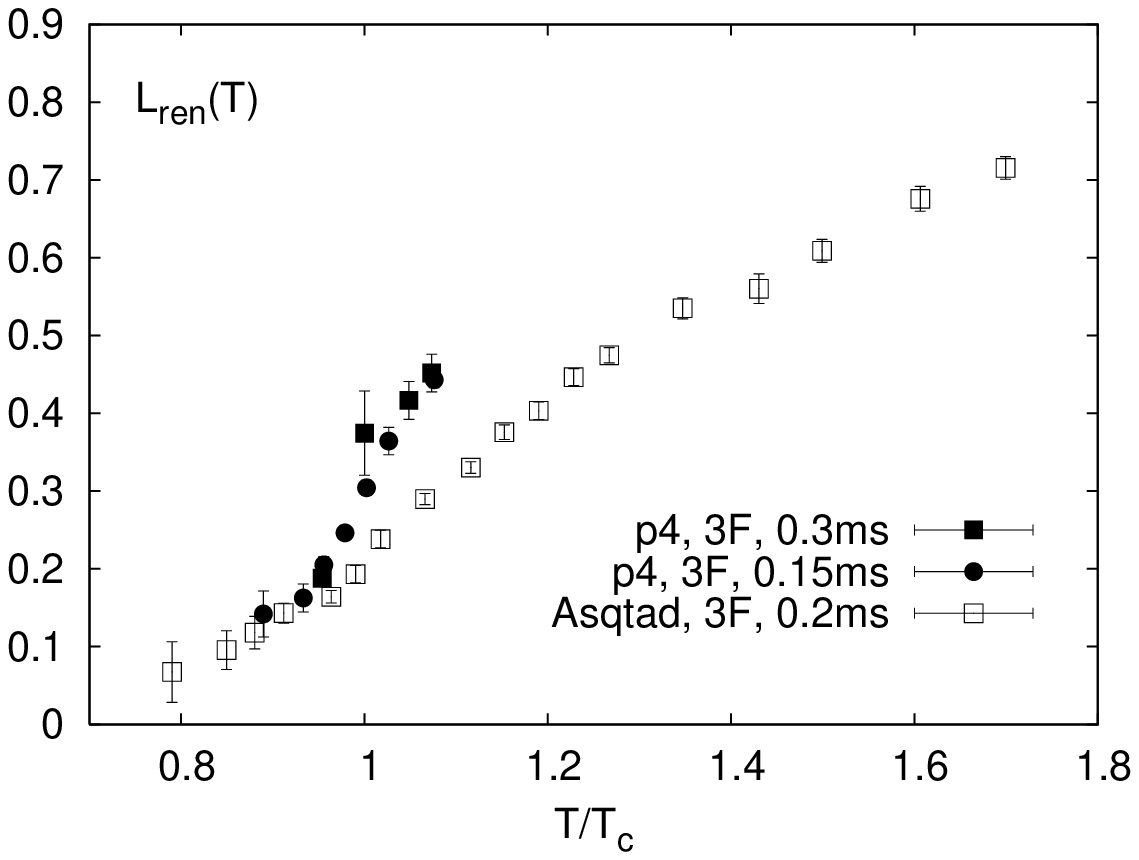}
\caption{The comparison of the  renormalized Polyakov loop calculated with p4 and Asqtad actions on $N_t=4$ 
lattices (left) and $N_t=6$ lattices (right). For $N_t=4$ we also shown the results from 2 flavor p4 calculations.
}
\label{fig:lren_comp}
\end{figure}

It is interesting to compare our results for the renormalized Polyakov loop with the calculations for
three degenerate flavors performed with Asqtad action \cite{petrov04} as well as with the two flavor calculations
with p4 action at larger quark masses \cite{okacz05}. The calculations with Asqtad
action have been done on lattices $12^3 \times 4$ and $12^3 \times 6$. Critical temperatures are 194(15)MeV  for $N_t=6$ and 199(8)MeV for $N_t=4$, $m_q=0.2m_s$\cite{levkova}. This is consistent with out own estimates from the maximum of susceptibility of the renormalized Polyakov loop.
The two flavor calculations with p4 action were performed on $16^3 \times 4$ lattice and for quark mass of about
$1.54m_s$ with $m_s$ being the physical strange quark mass.
This comparison is shown in Fig. \ref{fig:lren_comp} where the renormalized Polyakov loops is plotted in a wider temperature range.
We see some discrepancy in the transition region between the results of our study and earlier Asqtad calculations. For $N_t=4$ it can easily be explained by the large error for the determination of the critical temperature - we can pick a temperature in the allowed region which makes curves lie almost on each other. The $N_t=6$ case shows more dramatic descrepancy and here, apart from even larger ambiguity for the critical temperature, we have to point out that Asqtad simulation are done on a much smaller volume.

Most noticable deviations from the results of previous calculations are seen 
at higher temperatures. These  deviations
may come from the fact that the parameterization of the zero temperature potential given
by Eq. (\ref{t0pot}) may not be appropriate for the small lattice spacings corresponding to high temperatures.
Also the parameterization of the non-perturbative beta functions used in the present analysis is based on the 
analysis of $r_0$  for gauge coupling $\beta=6/g^2$ in the range between 3.3 and 3.4. It may not be accurate 
for $\beta$ values corresponding to high temperatures. This also could yield a discrepancy in the
value of the renormalized Polyakov loop in the high temperature region.

\section{Conclusions}
We have calculated the singlet free energy of a static quark anti-quark pair in full QCD
in the region of the small quark masses. We have found that this quantity shows little
cut-off dependence and can be calculated reliably on relatively coarse lattices.
The renormalized Polyakov loop derived from the large distance limit of the singlet
free energy has been also calculated and compared with earlier results. We have found that
the flavor and quark mass dependence of the renormalized Polyakov loop can be absorbed
almost entirely in the flavor and quark mass dependence of the transition temperature.

\end{document}